\documentclass[letterpaper, 10 pt, conference]{ieeeconf}  

\IEEEoverridecommandlockouts                              

\usepackage{graphics} 
\usepackage{epsfig} 
\usepackage{mathptmx} 
\usepackage{times} 
\usepackage{amsmath} 
\usepackage{amssymb}  
\usepackage{subcaption}
\usepackage{booktabs}
\usepackage{tabularx} 
\usepackage{multirow}
\usepackage{graphicx}
\usepackage{array}

\title{\LARGE \bf
Task Matters: Investigating Human Questioning Behavior in Different Household Service for Learning by Asking Robots}

\author{
Yuanda Hu$^{1}$, Jiani Hou$^{1}$, Junyu Zhang$^{1}$, Yate Ge$^{1}$, Xiaohua Sun$^{2}$, Weiwei Guo$^{1*}$%
\thanks{$^{1}$College of Design and Innovation, Tongji University, Shanghai, China. \{ydhu, 2231959, 2252482, geyate, weiweiguo\}@tongji.edu.cn}%
\thanks{$^{2}$School of Design, Southern University of Science and Technology, Shenzhen, China. Email: sunxh@sustech.edu.cn}%
\thanks{*Corresponding author: weiweiguo@tongji.edu.cn}%
}

\begin{document}

\maketitle
\thispagestyle{empty}
\pagestyle{empty}

\begin{abstract}


Learning by Asking (LBA) enables robots to identify knowledge gaps during task execution and acquire the missing information by asking targeted questions. However, different tasks often require different types of questions, and how to adapt questioning strategies accordingly remains underexplored. This paper investigates human questioning behavior in two representative household service tasks: a Goal-Oriented task (refrigerator organization) and a Process-Oriented task (cocktail mixing). Through a human-human study involving 28 participants, we analyze the questions asked using a structured framework that encodes each question along three dimensions: acquired knowledge, cognitive process, and question form. Our results reveal that participants adapt both question types and their temporal ordering based on task structure. Goal-Oriented tasks elicited early inquiries about user preferences, while Process-Oriented tasks led to ongoing, parallel questioning of procedural steps and preferences. These findings offer actionable insights for developing task-sensitive questioning strategies in LBA-enabled robots for more effective and personalized human-robot collaboration.

\end{abstract}

\section{Introduction}
Active learning has become an increasingly influential paradigm in robotics, enabling robots to iteratively query human users (oracles) for labels on informative samples during human-robot interaction. This process reduces uncertainty by enabling the robot to selectively acquire information about ambiguous or unfamiliar situations through human input~\cite{taylor2021active}. Among the various active learning approaches, Learning by Asking (LBA) has emerged as a particularly promising method, allowing robots to strategically acquire task-relevant knowledge by formulating targeted questions~\cite{misra2018learning, biyik2019asking, ren2023robots}. For example, Ren~\textit{et al.}~\cite{ren2023robots} approached the problem from an uncertainty modeling perspective, enabling robots to detect low-confidence situations and request help when needed, while Dai~\textit{et al.}~\cite{dai2024think} showed that robots performing object navigation tasks can enhance performance by asking questions to obtain additional information. Collectively, these studies underscore the potential of question-asking as an effective mechanism for improving robotic task performance.

\begin{figure}[htbp]
\centering
\includegraphics[width=\columnwidth]{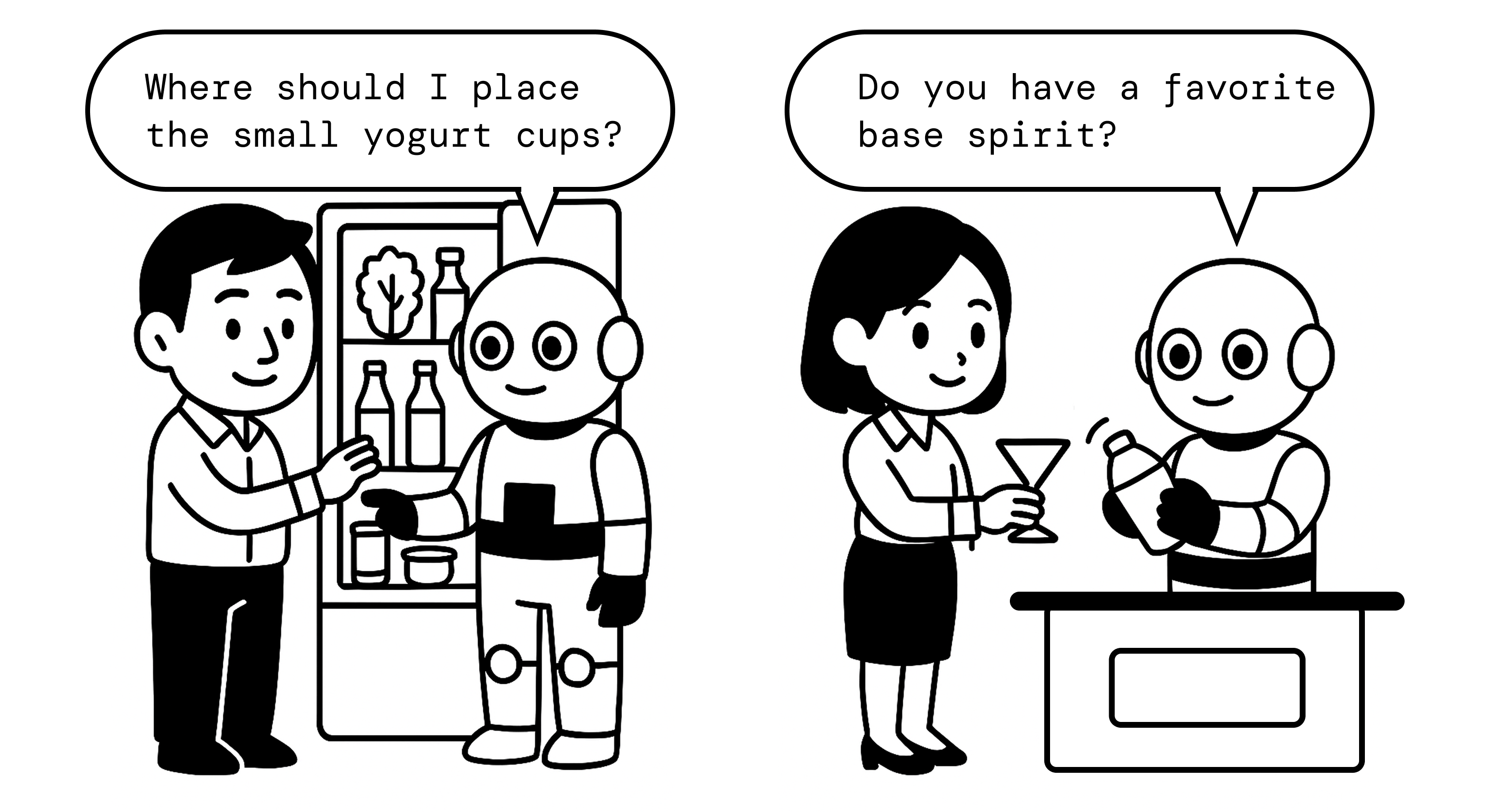}
\caption{Illustration of Learning by Asking in two household service tasks. Robots ask different types of questions depending on the nature of the task: Goal-Oriented (e.g., organizing a refrigerator, left) vs. Process-Oriented(e.g., mixing drinks, right).}
\label{fig:teasure}
\end{figure}

Despite these advances, identifying ``what is a good question'' remains a challenge. Previous research in human-robot interaction has addressed this issue from multiple perspectives~\cite{cakmak2012designing, shin2023uncertainty, shin2024caus, habibian2022here}. For instance, Cakmak~\textit{et al.}~\cite{cakmak2012designing} identified three types of questions that robots can use when learning new skills. Shin~\textit{et al.}~\cite{shin2023uncertainty} introduced a framework for generating more human-like questions when faced with uncertainty. While these studies offer valuable insights, they often do not address how such questioning strategies apply to specific real-world domains. In household service scenarios, where robots must frequently align with user preferences, question-asking plays a key role in acquiring personalized information. However, it remains unclear how questioning strategies should adapt to different tasks. 

To address this problem, we examine human questioning behavior in household service contexts. Specifically, we investigate whether people ask different types of questions when performing tasks with different procedural structures. We categorize household tasks into two types—\textit{Goal-Oriented} and \textit{Process-Oriented}—and select one representative task for each: refrigerator organization and cocktail mixing. To analyze the differences in questioning behavior, we conducted a human-human study and recorded participants' interactions. We then coded the questions and performed statistical analysis to uncover both the distributional patterns of question types and the higher-level questioning strategies used across tasks.

Our results reveal significant differences in both the types of questions asked and the temporal ordering of those questions across Goal-Oriented and Process-Oriented tasks. We also identify distinct questioning strategies employed by participants. These insights offer guidance for designing robots that can learn through question-asking in a task-sensitive manner.

Our contributions are threefold:
\begin{itemize}
    \item We conduct a human-human study in the context of household service, examining how questioning behavior differs between Goal-Oriented and Process-Oriented tasks.
    \item We provide a structured analysis of the categories and temporal patterns of questions using an established classification framework.
    \item We identify task-sensitive questioning strategies and offer empirical insights to guide the design of Learning by Asking (LBA) robots in household environments.
\end{itemize}

\section{Background}
Learning by Asking (LBA) represents an effective paradigm that allows robots to strategically gather information through questioning. This approach enhances robotic adaptability in changing environments and diverse task scenarios \cite{misra2018learning, biyik2019asking, ren2023robots}.
From a robot learning perspective, Valipour et al. \cite{valipour2017incremental} demonstrated that LBA increases task performance when robots face knowledge gaps or ambiguous situations. Additionally, this approach enables robots to better understand user intentions and adjust their behavior accordingly. Service robots can obtain personalized household preferences through specific questions, thus expanding their contextual knowledge base \cite{ayub2022don}. From the user experience perspective, providing feedback about information obtained from human responses improves teaching accuracy and efficiency \cite{cakmak2012designing}. Research by Doering et al. \cite{doering2019curiosity} revealed that curiosity-driven questioning systems significantly enhance both learning efficiency and interaction quality. Furthermore, in temporal or procedural task scenarios, LBA strategies have demonstrated superior effectiveness in assisting non-expert users \cite{racca2018active}.

A key challenge in LBA research is defining ``good questions''—those that demonstrate both contextual relevance \cite{mi2024convqg} and informational value \cite{rothe2017question}. Recent human-robot interaction (HRI) studies have advanced both theoretical frameworks and methods for questioning, improving robots' ability to learn in complex interactive settings. Shin \cite{shin2023uncertainty} introduced a multidimensional framework enabling robots to formulate questions under uncertainty, effectively reducing task ambiguities. The CAUS dataset established a human cognition-based question generation mechanism designed to train robots in generating "meaningful questions" \cite{shin2024caus}. Building on these research, Habibian et al. \cite{habibian2022here} combined human and robot perspectives, allowing robots to ask questions while showing their learning state to human partners, which promotes shared understanding. While these studies establish theoretical foundations for question formulation, limited research exists on applying these concepts to specific tasks. Our work fills this gap by analyzing question types and sequencing patterns across different tasks.

Analyzing human behavior to inspire and guide robot design is a common research approach. For language-based interactions, Cakmak et al. \cite{cakmak2012designing} studied how humans ask questions during teaching to identify key question categories, providing design guidelines for robots to learn from demonstration (LfD). Similarly, Narvekar et al. \cite{narvekar2020curriculum} examined human teaching sequences to develop models aligned with curriculum learning principles.  In non-verbal interaction research, Schmidt et al. \cite{schmidt2024investigating} investigated how humans use gestures to communicate intent in complex collaborative settings, informing more effective robot movement design. Huang and Mutlu \cite{huang2016anticipatory} examined human gaze patterns during collaboration and applied these findings to improve robot performance in object handover tasks. Following this approach, our study investigates human questioning behavior in specific household tasks to identify patterns that can enhance the development of Learning by Asking (LBA) strategies for service robots.

\section{Methodology}
\subsection{Classification of Task}
Delivering high-quality user experiences with household service robots depends on their ability to perform tasks effectively. While previous research has addressed robot spatial awareness \cite{dai2024think} and dynamic response capabilities \cite{racca2018active}, the relationship between task structure and questioning strategies remains underexplored. To investigate this relationship, we first categorized household service robot tasks by reviewing relevant literature \cite{inproceedings1,6483517,10.5555/2035304.2035336,6218159} and identifying common task types. Based on this analysis, we classified household tasks into two categories: \textit{Goal-Oriented} and \textit{Process-Oriented}. Goal-Oriented tasks focus on achieving a final state with flexible execution order (e.g., organizing a refrigerator or sorting a wardrobe), while Process-Oriented tasks require following specific sequential steps to reach the desired outcome (e.g., cocktail mixing or preparing dinner).

\subsection{Study Design}
\begin{figure}[htbp]
    \centering
    \includegraphics[width=\columnwidth]{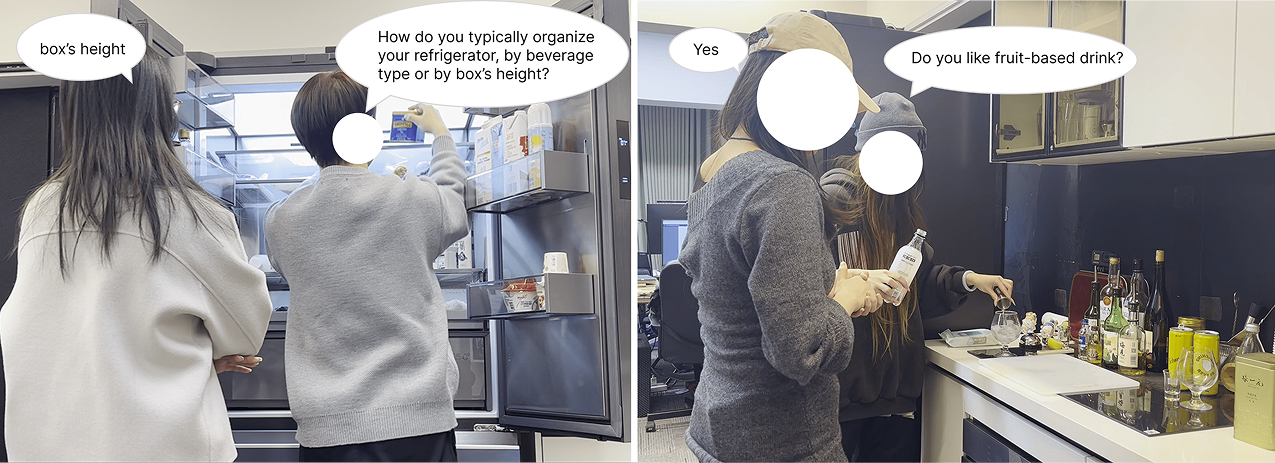}
    \caption{Participants in the human-human study performing two representative household tasks: refrigerator organization (left) and cocktail mixing (right).}
    \label{fig:task_examples}
\end{figure}

To explore how humans ask questions in the Goal-Oriented task and the Process-Oriented task, we selected refrigerator organization (\textbf{T1}) and cocktail mixing (\textbf{T2}) as representative tasks and conducted a human-human study to analyze the questions asked by the participants, as shown in Figure~\ref{fig:task_examples}.

The study required two participants to collaboratively complete both tasks. We first recruited 40 paid participants from the university community through questionnaires, where they provided individual information and rated their familiarity with both tasks. We then paired participants who were familiar with both tasks with participants who were not familiar with both tasks. Ultimately, 14 pairs (28 participants in total) were invited to participate in the study, with an average participant age of 22 years.

Before each session, participants were assigned roles: those familiar with the tasks acted as teachers, while those unfamiliar acted as learners. Learners were instructed to complete each task according to the teacher’s habits and preferences, but they could only obtain this information by asking questions. Prior to starting, teachers were given time to familiarize themselves with the environment, envisioning it as their own home and considering their typical habits and preferences for completing the tasks. Following this preparation phase, the experiment commenced with participants first performing the refrigerator organization task, followed by the cocktail mixing task. Task completion criteria were determined by the participants. All sessions were recorded using a GoPro Hero 9 camera to facilitate subsequent analysis.

\subsection{Question Type Classification}
To analyze the questions asked by participants in our study, we used the framework proposed by Shin \textit{et al.}~\cite{shin2023uncertainty}, which categorizes questions into three types: \textit{K-type} (missing knowledge),\textit{ C-type} (cognitive process), and \textit{Q-type} (question expression), as shown in Table~\ref{tab:type}.  The K-type refers to the specific knowledge gap the question aims to fill, such as object identity, quantity, spatial layout, task procedure, or user preference. The C-type reflects the cognitive process needed to answer the question, including retrieval, comprehension, application, or evaluation, indicating the level of mental effort involved. The Q-type captures how the question is formulated, such as confirmation, definition, quantification, or judgment, highlighting its sentence form of question.

Using this classification, we gained a deeper understanding of both the content and structure of the questions in the participant dialogues.

\subsection{Research Questions}
This human-human study investigates how participants ask questions while performing different tasks. The collected data serve as a foundation for designing robots that learn by asking. We address the following research questions:

\begin{itemize}
    \item \textbf{RQ1}: Do participants use all types of question when performing Goal-Oriented and Process-Oriented tasks?
    \item \textbf{RQ2}: Do participants ask different types of question when performing Goal-Oriented and Process-Oriented tasks?
    \begin{itemize}
        \item \textbf{H1}: In both tasks, the knowledge gained by asking questions differed.
        \item \textbf{H2}:In both tasks, the cognitive process of participants are different when asking questions.
        \item \textbf{H3}: In both tasks, the expression of questions are different.
    \end{itemize}
    \item \textbf{RQ3}: Is there a clear strategy in the order of questions participants ask while completing a task? Furthermore, do they use different questioning strategies depending on the type of task?
\end{itemize}


\section{Results}

\begin{table}[]
\caption{K-type, C-type, and Q-type Categories}
\label{tab:type}
\renewcommand{\arraystretch}{1.4}
\resizebox{\columnwidth}{!}{%
\begin{tabular}{llllll}
\hline
                                              & \textbf{\#}  & \textbf{Categories}                             &                                               & \textbf{\#}  & \textbf{Content}               \\ \hline
\multicolumn{1}{l|}{\multirow{11}{*}{K-type}} & K1  & \multicolumn{1}{l|}{Identity}          & \multicolumn{1}{l|}{\multirow{15}{*}{Q-type}} & Q1  & Verification          \\
\multicolumn{1}{l|}{}                         & K2  & \multicolumn{1}{l|}{Class}             & \multicolumn{1}{l|}{}                         & Q2  & Case specification    \\
\multicolumn{1}{l|}{}                         & K3  & \multicolumn{1}{l|}{Attributes}        & \multicolumn{1}{l|}{}                         & Q3  & Concept completion    \\
\multicolumn{1}{l|}{}                         & K4  & \multicolumn{1}{l|}{Quantities}        & \multicolumn{1}{l|}{}                         & Q4  & Feature specification \\
\multicolumn{1}{l|}{}                         & K5  & \multicolumn{1}{l|}{Spatial layout}    & \multicolumn{1}{l|}{}                         & Q5  & Quantification        \\
\multicolumn{1}{l|}{}                         & K6  & \multicolumn{1}{l|}{Temporal relation} & \multicolumn{1}{l|}{}                         & Q6  & Definition            \\
\multicolumn{1}{l|}{}                         & K7  & \multicolumn{1}{l|}{Contents}          & \multicolumn{1}{l|}{}                         & Q7  & Comparison            \\
\multicolumn{1}{l|}{}                         & K8  & \multicolumn{1}{l|}{Procedure}         & \multicolumn{1}{l|}{}                         & Q8  & Interpretation        \\
\multicolumn{1}{l|}{}                         & K9  & \multicolumn{1}{l|}{Causality}         & \multicolumn{1}{l|}{}                         & Q9  & Cause elucidation     \\
\multicolumn{1}{l|}{}                         & K10 & \multicolumn{1}{l|}{Intention}         & \multicolumn{1}{l|}{}                         & Q10 & Result account        \\
\multicolumn{1}{l|}{}                         & K11 & \multicolumn{1}{l|}{Internal state}    & \multicolumn{1}{l|}{}                         & Q11 & Intention disclosure  \\ \cline{1-3}
\multicolumn{1}{l|}{\multirow{4}{*}{C-type}}  & C1  & \multicolumn{1}{l|}{Knowledge}         & \multicolumn{1}{l|}{}                         & Q12 & Method explication    \\
\multicolumn{1}{l|}{}                         & C1  & \multicolumn{1}{l|}{Comprehension}     & \multicolumn{1}{l|}{}                         & Q13 & Expectation           \\
\multicolumn{1}{l|}{}                         & C1  & \multicolumn{1}{l|}{Operation}         & \multicolumn{1}{l|}{}                         & Q14 & Judging               \\
\multicolumn{1}{l|}{}                         & C1  & \multicolumn{1}{l|}{Evaluation}        & \multicolumn{1}{l|}{}                         &     &                       \\ \hline
\end{tabular}%
}
\end{table}

To understand the questions posed during the conversations, we transcribed the video data and extracted the questions. Then, we encoded these questions using the framework proposed by Shin et al. \cite{shin2023uncertainty}, which provides a structured method for categorizing questions from the perspective of human information acquisition  (detailed in the \textit{Methodology} section). 

\subsection{Usage of Question Types}
\begin{figure}[htbp]
\centering
\includegraphics[width=\columnwidth]{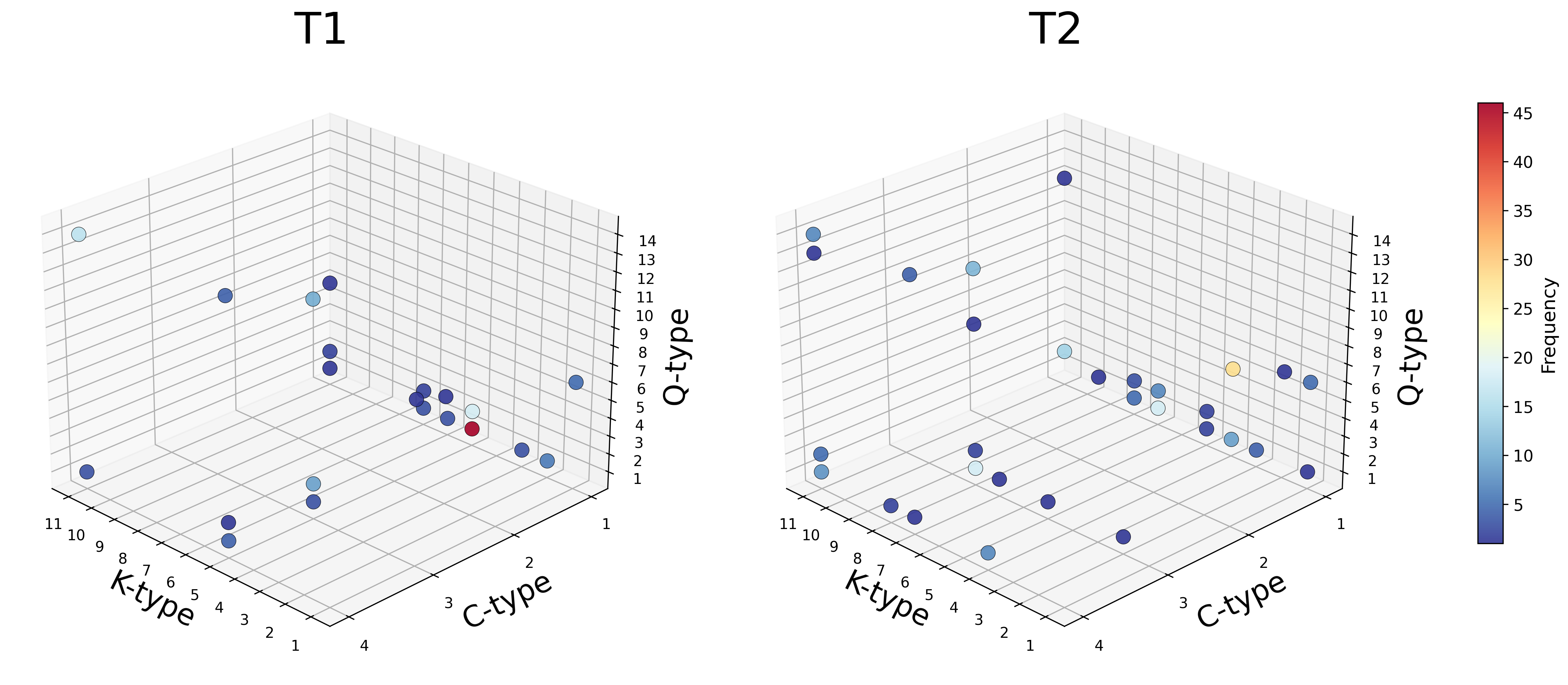}
\caption{3D visualization of question type combinations in \((K, C, Q)\) space. Each point represents a unique combination used during task execution. Color encode frequency.}
\label{fig:sparse}
\end{figure}

To address \textbf{RQ1}, we analyzed the usage of the question space by visualizing the distribution of K-type, C-type, and Q-type combinations in a 3D space (Figure~\ref{fig:sparse}). Each point in the space represents a unique \((K, C, Q)\) tuple, with its color indicating the frequency of occurrence. The theoretical space contains \(11 \times 4 \times 14 = 616\) possible combinations. However, the observed data show that only 21 combinations were used in T1 and 30 in T2, corresponding to sparsity levels of 96.59\% and 95.13\%, respectively.

This high sparsity indicates that humans are highly selective when generating questions for a specific task. In T1, the top combinations are K5–C1–Q1, K5–C3–Q2, and K11–C4–Q14. These reflect verification of spatial placement, operational requests, and evaluation of user preferences. For example, participants asked “Should I put the milk on the top shelf?” (K5, C1, Q1), “Where should the yogurt go?” (K5, C3, Q2), and “Would you prefer vegetables on the left side?” (K11, C4, Q14). 
In T2, the top combinations are K4–C1–Q5, K7–C1–Q6, and K8–C1–Q6. These reflect factual queries about amounts, ingredient composition, and procedural steps. For example, participants asked “How much juice should I pour?” (K4, C1, Q5), “What goes into this cocktail?” (K7, C1, Q6), and “What is the mixing order?” (K8, C1, Q6).


\subsection{Distributional Differences in Question Types}
To address \textbf{RQ2}, we examined whether participants asked different types of questions when performing T1 and T2. Specifically, we conducted Fisher's exact tests to determine whether there were significant differences in the distributions of knowledge category (K-type), cognitive process (C-type), and question expression (Q-type) between the two tasks. These analyses correspond to hypotheses \textbf{H1}, \textbf{H2}, and \textbf{H3}, respectively.

\begin{figure}[b]
\centering
\includegraphics[width=\columnwidth]{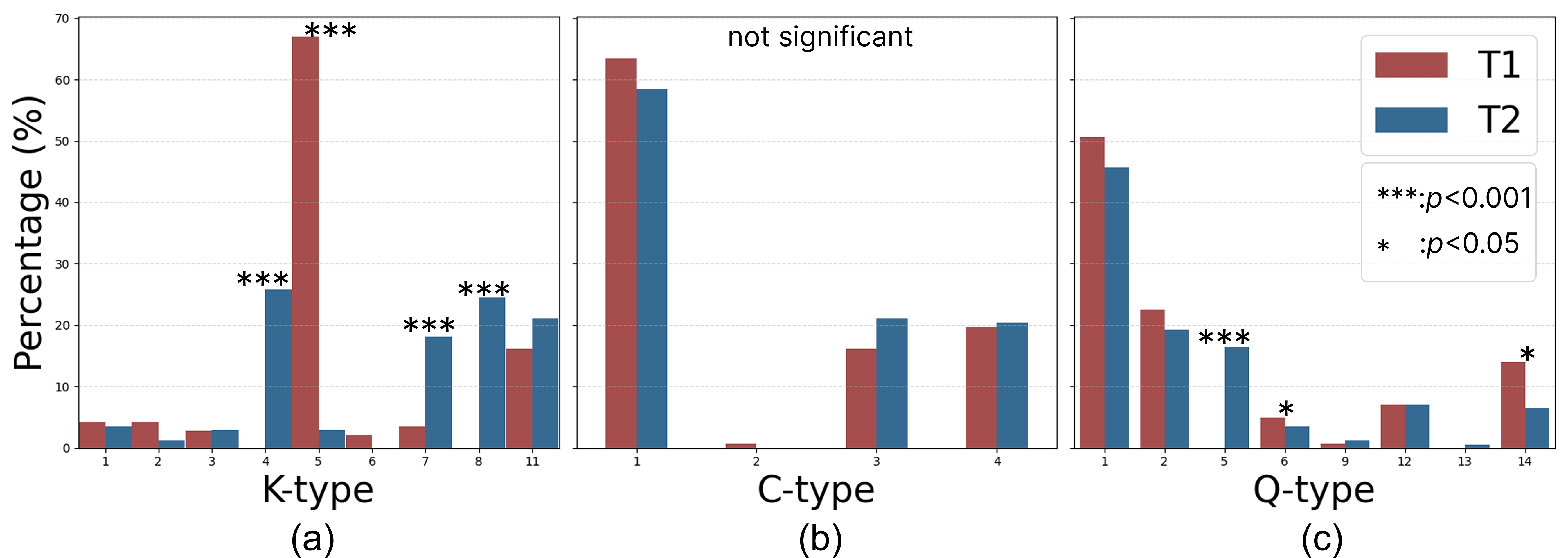}
\caption{Distributions of K-type, C-type, and Q-type questions across T1 and T2.}
\label{fig:bar}
\end{figure}

\textbf{H1 (K-type):}  As shown in Figure~\ref{fig:bar} (a), in the K-type dimension, participants in T1 predominantly asked K5 (spatial arrangement) questions, followed by K11 (internal state), with relatively few questions in other categories. In contrast, T2 participants focused more on K4 (object quantity), K7 (content), K8 (task procedure), and K11. Fisher’s exact test revealed significant differences for K4, K5, K7, and K8 ($p < .001$), indicating that participants sought different types of knowledge depending on the task. 

\textbf{H2 (C-type):} As shown in Figure~\ref{fig:bar} (b), in the C-type dimension, both T1 and T2 participants primarily relied on C1 (knowledge retrieval). Fisher’s exact test found no significant differences across question categories ($p > .05$), suggesting that the cognitive processes underlying question-asking were consistent across tasks. Thus, H2 is not supported.  

\textbf{H3 (Q-type):} As shown in Figure~\ref{fig:bar} (c), in the Q-type dimension, T2 participants exhibited a strong preference for Q5 (quantification), whereas T1 participants asked more Q1 (confirmation), Q6 (definition), and Q14 (judgment) questions. Fisher’s exact test confirmed significant differences in Q5 ($p < .001$), Q6 ($p < .05$), and Q14 ($p < .05$), supporting H3.  

These results reveal that participants’ question types varied systematically with tasks. T1 participants focused on spatial reasoning, conceptual clarification, and subjective interpretation, while T2 participants emphasized quantitative and procedural reasoning. Task type significantly influenced both the knowledge sought and question phrasing. However, it did not affect the underlying cognitive processes.

\subsection{Temporal Patterns and Questioning Strategies}

\begin{table*}[htbp]
\centering
\caption{Overview of the Three Questioning Strategies}
\label{tab:strategies}
\renewcommand{\arraystretch}{1.8}

\begin{tabularx}{\textwidth}{l
>{\centering\arraybackslash}m{0.20\textwidth}
>{\arraybackslash}m{0.25\textwidth}
>{\arraybackslash}m{0.25\textwidth}}
\hline
\textbf{Strategy} & \textbf{Visualization} & \textbf{Description} & \textbf{Example} \\
\hline

\textbf{User-Preference-First} & 
\includegraphics[width=0.22\textwidth]{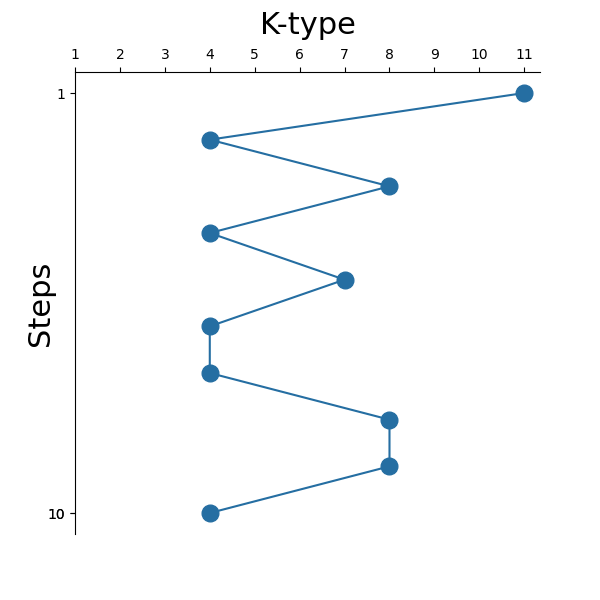} & 
The participant begins the questioning sequence by inquiring about the user's personal habits or preferences. This high-level information serves as a foundation for planning, allowing subsequent task-related decisions to align with the user’s expectations and routines. &
\textit{“Do you have a preferred way of grouping items in the fridge?” (K11)}\newline
$\rightarrow$ \textit{“Should drinks go on the door shelf or inside?” (K11)}\newline
$\rightarrow$ \textit{“Where would you usually place dairy products?” (K5)}\newline
$\rightarrow$ \textit{...}\newline
$\rightarrow$ \textit{“Is this how you normally organize vegetables?” (K11)} \\
\hline

\textbf{Parallel Exploration} & 
\includegraphics[width=0.22\textwidth]{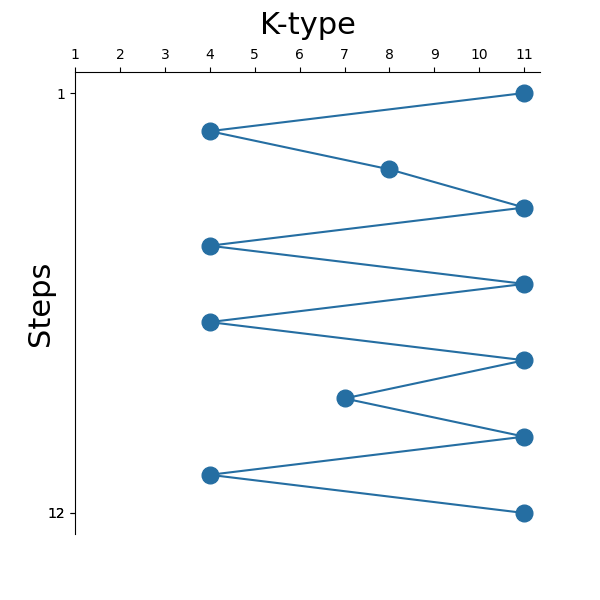} & 
The participant interleaves questions about user preferences or internal state with questions focused on specific task operations. This strategy reflects a dynamic process of integrating user needs with procedural decisions throughout task execution. &
\textit{“What flavor of cocktail do you prefer?” (K11)}\newline
$\rightarrow$ \textit{“How many ingredients should it contain?” (K4)}\newline
$\rightarrow$ \textit{“What is the mixing sequence?” (K8)}\newline
$\rightarrow$ \textit{“Do you like fruit-based mixers?” (K11)}\newline
$\rightarrow$ \textit{...}\newline
$\rightarrow$ \textit{“The drink is ready—would you like any adjustments?” (K11)} \\
\hline

\textbf{Direct Querying} & 
\includegraphics[width=0.22\textwidth]{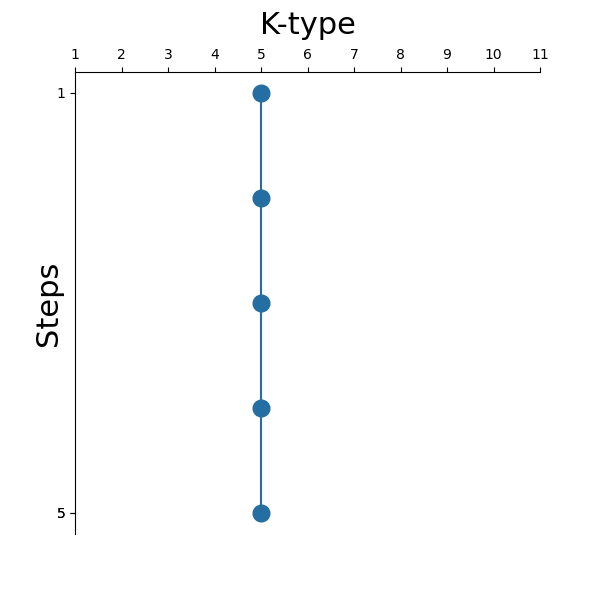} & 
The participant directs all questions toward specific task instructions—such as object placement or execution steps—without proactively considering the user's personal preferences. This strategy reflects a reactive approach in which the participant follows surface-level directives without engaging in deeper reasoning about the user’s underlying habits or needs. &
\textit{“The milk won’t fit—should I put it on the lower shelf?” (K5)}\newline
$\rightarrow$ \textit{“Where should I place the small yogurt cups?” (K5)}\newline
$\rightarrow$ \textit{“Where do the fresh vegetables go?” (K5)}\newline
$\rightarrow$ \textit{“What should I do if the left side is full?” (K5)}\newline
$\rightarrow$ \textit{“Is it okay to place canned food on the upper right shelf?” (K5)} \\
\hline
\end{tabularx}
\end{table*}

To address \textbf{RQ3}, we focus on the K-type dimension, as the K-type dimension reflects the type of information extracted from each question. 

We first visualized the question sequences from T1 and T2, as shown in \textit{visualization} column in Table~\ref{tab:strategies}. In these visualizations, the x-axis represents different K-type categories, while the y-axis indicates the chronological order of the questions asked during the execution of task. A total of 26 visualizations were generated. Next, we invited two researchers familiar with dialogue theory to classify these visualizations. After two rounds of discussion, they reached a consensus and categorized the questioning strategies into three types, as summarized in Table~\ref{tab:strategies}.  

Specifically, we first examined the overall shape of the visualizations. Based on their shapes, we identified two broad categories: disorganized sequences, where multiple K-type categories were used, and structured sequences, where participants relied primarily on one or two K-type categories. Then, we conducted a more detailed analysis of K-type usage. Since one key advantage of learning by asking is the ability to quickly infer user preferences, we selected K11 (internal state) as a key classification criterion. And classified according to the following methods:

\begin{itemize}
    \item If the first question in a sequence belonged to K11 and subsequent questions mainly focused on other categories, the sequence was classified as following the \textit{User-Preference-First Strategy}. This indicates that participants prioritized gathering user preference information before executing the task.  
    \item If K11 appeared consistently throughout the sequence or was present in both the early and later stages, the sequence was categorized as the \textit{Parallel Exploration Strategy}. This suggests that participants continuously integrated user preferences with operational information during task execution.  
    \item Finally, if K11 was completely absent, the sequence was classified as the \textit{Direct Querying Strategy}, reflecting a focus solely on acquiring task-related operational information without considering user-specific preferences.  
\end{itemize}

\begin{figure}[htbp]
\centering
\includegraphics[width=0.8\columnwidth]{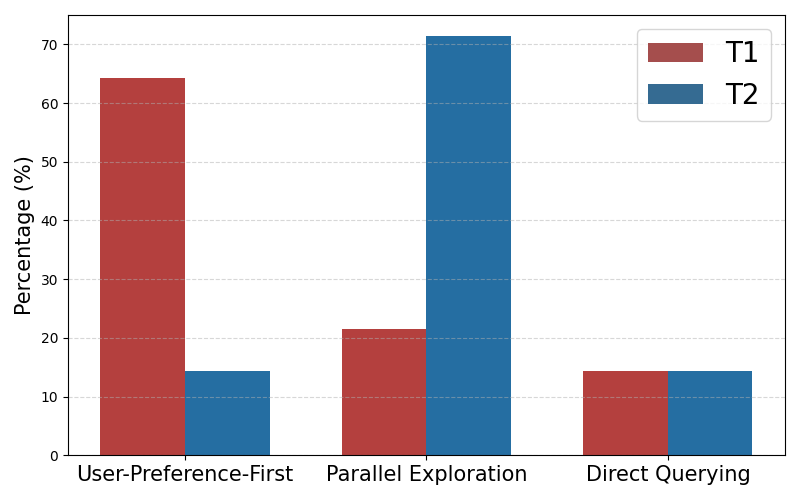}
\caption{Distribution of questioning strategies across task types.}
\label{fig:bar2}
\end{figure}

In the end, we categorized the questioning strategies used in T1 and T2 tasks. As shown in Figure~\ref{fig:bar2}, for the 14 T1 tasks, the \textit{user-preference-first} strategy was adopted in 9 cases (64.3\%), the \textit{parallel exploration} strategy in 3 cases (21.4\%), and the \textit{direct inquiry} strategy in 2 cases (14.3\%). In contrast, for the 14 T2 tasks, only 2 cases (14.3\%) followed the \textit{user-preference-first} strategy, while 10 cases (71.4\%) employed the \textit{parallel exploration} strategy, and 2 cases (14.3\%) used the \textit{direct querying} strategy.  
Overall, participants in T1 frequently adopted a \textit{user-preference-first} strategy, whereas those in T2 primarily employed a \textit{parallel exploration} strategy.

\section{Discussion}
\subsection{Task-Specific Patterns in Question Content}

Our study found that participants do not use all types of questions when completing a task. Instead, they tend to ask questions that are most relevant to achieving their goal. For example, in the refrigerator organization task (T1), most questions focused on K5 (spatial layout) and K11 (user preferences). In the cocktail-making task (T2), the main focus was on K4 (quantities), K7 (contents), and K8 (procedure). This suggests that participants selectively ask questions to obtain the most relevant information, helping them complete the task more efficiently. Therefore, when designing task-oriented questioning strategies, it is important to consider the specific characteristics of the task to ensure that questions effectively support task completion. For example, when designing prompts for process-oriented tasks\cite{ren2023robots}, it may be helpful to first explain the task itself, then highlight the impact of different actions, and guide the robot to focus on key steps. This approach helps the robot obtain the most necessary information instead of asking questions randomly.  

Additionally, we found that while participants prioritized different types of knowledge and used different question phrasing in different tasks, their overall thinking strategies remained similar. This may be because completing a task is a complex process that involves repeatedly gathering information about objects, intentions, and events\cite{shin2024caus}, rather than just retrieving items or following instructions. Therefore, the context of the question is crucial, as it directly influences the robot’s reasoning process. For example, if a robot has already asked about the purpose of an object, it should remember this information and later inquire about its intended use within the task. Furthermore, we recommend optimizing questioning strategies based on the type of task. For instance, in process-oriented tasks, increasing the proportion of process-related questions can help the robot acquire more essential knowledge, ultimately improving its efficiency in completing the task.

\subsection{Task-Specific Patterns in Question Sequencing}
In everyday collaborative tasks—such as cooking together or organizing shared spaces—human partners typically rely on timely questions to clarify goals, understand preferences, and coordinate actions. An effective and well-crafted questioning strategy plays a crucial role in successful collaboration, especially when tasks require integrating background knowledge and user-specific needs. 

Participants employed different questioning strategies in Goal-Oriented and Process-Oriented scenarios. In the Goal-Oriented task, participants commonly initiated interactions by first eliciting user preferences, forming a User-Preference-First strategy. This approach helped establish high-level constraints early, guiding decisions about object placement. Such behavior suggests a proactive effort to reduce cognitive uncertainty by narrowing the decision space before engaging with the task environment. In contrast, participants in the Process-Oriented task adopted a Parallel Exploration strategy, alternating between procedural and preference-related questions. This pattern reflects the incremental nature of procedural tasks, which often require step-by-step adjustments and benefit from real-time integration of user input. The reactive Direct Querying strategy—focused on surface-level operations without considering user intent—was rarely observed. Its low prevalence highlights a general preference among participants to engage in goal-aware, personalized inquiry rather than relying on fixed action sequences.

These findings suggest that effective robot assistants should adjust their question-asking strategies according to task structure. For tasks involving open-ended goals, robots should prioritize early clarification of user preferences to build a guiding model of task constraints. For stepwise procedural tasks, they should support ongoing information gathering and adapt to user input during execution. Recent work by Patel and Chernova~\cite{patel2025robot} reinforces this view, showing that robot assistants can improve performance by inferring user preferences from limited feedback. Our results complement this direction by demonstrating how human partners naturally engage in strategic, context-aware questioning to support collaborative success. Together, these insights emphasize the value of incorporating task-sensitive and user-aligned questioning strategies into robot learning systems, participants needed to continuously ask user preference questions throughout the task to adjust their actions accordingly.


These insights could inform current work on LBA systems such as those developed by Ren et al.~\cite{ren2023robots} and Dai et al.~\cite{dai2024think}, potentially enhancing their ability to acquire task-relevant information through more natural and effective questioning strategies. By implementing task-specific question templates based on our findings, these systems could better adapt their information-seeking behavior to different task contexts.

\section{Limitations and Future Work}
While our study provides meaningful insights into task-aware questioning behavior, it has several limitations. Although we explored both Goal-Oriented and Process-Oriented task categories, each was represented by only a single task: refrigerator organization and cocktail mixing. These tasks may not capture the full range of variation within their respective categories. Therefore, our findings may not fully apply to other tasks with different characteristics. In addition, while the study reveals when and what types of questions participants tend to ask, it does not address how effective these questions are in terms of information value. Prior research has shown that people often struggle to formulate optimally informative questions~\cite{rothe2017question}, indicating that even the dominant strategies observed here may offer limited information gain. Another limitation is the lack of empirical validation through robot implementation. The identified strategies have not yet been tested in robotic systems, making their practical effectiveness uncertain.

Future research could deploy these strategies in real or simulated environments to assess the utility of different question types and sequences in achieving task objectives, as well as evaluate their impact on user experience. Furthermore, developing computational models that dynamically select or adapt questioning strategies based on task context would support integration into existing LBA frameworks.

\section{CONCLUSION}
This paper explored how task structure influences human questioning behavior in household service scenarios, with the aim of informing the design of Learning by Asking (LBA) robots. We categorized tasks into Goal-Oriented and Process-Oriented types and analyzed the questions asked in each. This revealed clear differences in question types, frequencies, and sequencing strategies. Our findings demonstrate that humans adapt their questioning patterns based on task demands, favoring preference-first strategies in Goal-Oriented tasks and parallel exploration in Process-Oriented tasks. These insights offer concrete guidance for the design of context-sensitive, user-aligned questioning strategies in LBA-enabled service robots. Future work will extend this framework to a broader set of task domains and evaluate the effectiveness of these strategies in real-world robotic deployments.

\addtolength{\textheight}{-10cm}   
\bibliographystyle{IEEEtran}
\bibliography{ref}
\end{document}